\documentstyle[aps,multicol]{revtex}

\begin{document}
\draft

\title{Reduction of optimum light power with 
Heisenberg-limited photon-counting noise in interferometric 
gravitational-wave detectors}
\author{Constantin Brif\thanks{E-mail: cbrif@ligo.caltech.edu}}
\address{LIGO Project, California Institute of Technology,
Pasadena, CA 91125}
\maketitle

\begin{abstract}
We study how the behavior of quantum noise, presenting the 
fundamental limit on the sensitivity of interferometric 
gravitational-wave detectors, depends on properties of
input states of light. We analyze the situation with specially
prepared nonclassical input states which reduce the 
photon-counting noise to the Heisenberg limit. This results
in a great reduction of the optimum light power needed to 
achieve the standard quantum limit, compared to the usual 
configuration. 
\end{abstract}


\begin{multicols}{2}

Since the pioneering work by Caves \cite{Caves80,Caves81}, it 
is well understood that two sources of quantum noise---the 
photon-counting noise and the radiation-pressure 
noise---constitute the fundamental limitation on the sensitivity 
of an interferometric gravitational-wave detector. 
These limitations will be of potential importance in long-baseline 
interferometric detectors which are currently under construction 
(the LIGO project \cite{Abram92,Black97} in the United States and 
the French-Italian VIRGO project \cite{Caron97} in Europe are the 
largest ones). 
For example, the photon-counting shot noise will dominate at the 
gravitational-wave frequencies above 1 kHz in the VIRGO detector 
and above 200 Hz in the initial LIGO detector. With a further 
reduction of the thermal noise, planned in the advanced LIGO 
interferometer, the role of the shot noise will be even more 
important. 

For a coherent laser beam of light power $P$, the shot noise
associated with photon-counting statistics scales as $P^{-1/2}$
and the radiation-pressure noise scales as $P^{1/2}$. The 
contributions of these two sources of noise will be equal for 
some optimum value $P_{\mathrm{opt}}$ of the light power.
Provided that classical sources of noise (such as thermal and 
seismic) are sufficiently suppressed, the interferometer with the
optimum light power will work at the so-called standard quantum 
limit (SQL). A simple quantum calculation, based on the use of the
Heisenberg uncertainty principle, gives the SQL for the measurement
of the relative shift $z=z_2 - z_1$ in the positions of two end
mirrors:
\begin{equation}
  \label{eq:zSQL}
(\Delta z)_{\mathrm{SQL}} = \sqrt{2 \hbar \tau / m} ,
\end{equation}
where $m$ is the mass of each end mirror and $\tau$ is the 
measurement time. For modern long-baseline interferometers (like
LIGO and VIRGO), Fabry-Perot cavities are used in the arms; so
$\tau$ is actually the cavity storage time,
$\tau = L {\mathcal{F}} / \pi c$, where $L$ is the cavity length, 
${\mathcal{F}}$ is the finesse, and $c$ is the velocity of light. 
The optimum power, for which the SQL is achieved, is 
\begin{equation}
  \label{eq:Popt}
P_{\mathrm{opt}} = \frac{m L^2}{\omega \tau^4} ,
\end{equation}
where $\omega$ is the light angular frequency. For the initial
LIGO configuration, the mirror mass is $m \simeq 11\, \mathrm{kg}$, 
the cavity length is $L \simeq 4\, \mathrm{km}$, the finesse is
${\mathcal{F}} \simeq 200$, and the wavelength of the Nd:YAG laser is 
$\lambda \simeq 1.064\, \mu\mathrm{m}$ 
($\omega \simeq 1.77 \times 10^{15}\, \mathrm{Hz}$).
This gives the cavity storage time 
$\tau \simeq 8.5 \times 10^{-4}\, \mathrm{s}$ and an effective 
number of bounces $b = \tau c/ 2 L \simeq 32$. 
The corresponding optimum laser power is 
$P_{\mathrm{opt}} \simeq 191\, \mathrm{kW}$ 
and the SQL of the position shift measurement is 
$(\Delta z)_{\mathrm{SQL}} \simeq 1.24 \times 10^{-19}\, 
\mathrm{m}$. 
Achieving this SQL will make possible to measure 
gravitational waves with amplitudes $h$ greater than 
$\sim 3 \times 10^{-23}$.

Presently, the available laser light power is insufficient for 
achieving the SQL (for example, in the initial LIGO configuration
the input laser power is $6\, \mathrm{W}$ and the power recycling
gain is about $30$). Therefore, in advanced LIGO configurations, it 
is planned \cite{Black97} to reduce the shot noise by using more 
powerful lasers, in conjunction with the power-recycling technique
\cite{Drev83,Hello98}. 
However, for very high laser power, one encounters serious
technical problems related to nonuniform heating of the
cavity mirrors caused by absorption of even a small portion of 
circulating light. The resulting thermal aberrations can
seriously deteriorate the performance of the interferometer
\cite{Hello98}. Therefore, it will be very interesting to study 
possibilities for achieving the SQL with low light power. 

The gravitational-wave detection community is quite familiar 
with the intriguing idea by Caves \cite{Caves81} to reduce the 
photon-counting noise by squeezing the vacuum fluctuations at 
the unused input port. 
During the last decade, other interesting ideas has been 
developed in the field of theoretical quantum optics, based on 
the use of nonclassical photon states for the quantum noise 
reduction in idealized optical interferometers 
\cite{YMK86,HoBu93,SaMi95,HiMl93,BM96,Kim98}.
The main theoretical motivation of all those papers was to show 
the possibility of beating the shot-noise limit and achieving 
the fundamental Heisenberg limit for the photon-counting noise 
in an ideal interferometric measurement. 

The aim of the present work is to show that the optimum light
power needed for the SQL operation of an interferometric 
gravitational-wave detector with movable mirrors can be greatly
reduced by the use of nonclassical states of light with the
Heisenberg-limited photon-counting noise. This result means 
that Heisenberg-limited interferometry is not only interesting
for a demonstration of the fundamental uncertainty principle,
but can be also important for the experimental detection
of gravitational waves.

Let us consider a long-baseline Michelson interferometer whose 
arms are equipped with high-finesse Fabry-Perot cavities, with
end mirrors serving as free test masses. 
In the quantum description, two modes of the light field enter
the interferometer through the two input ports of a 50-50 beam
splitter. After being mixed in the beam splitter, the light modes
spent time $\tau$ in the Fabry-Perot cavities, and then leave
the interferometer (through the same beam splitter, but in the
opposite direction). The photons leaving the interferometer in
the output modes are counted by two photodetectors. 
A gravitational wave incident on the interferometer will cause
a relative shift $z=z_2 - z_1$ in the positions of two end
mirrors, which results in the phase shift 
$\phi = (\omega \tau/ L) z$
between the two arms.

The performance of such an interferometer can be analyzed in the
Heisenberg picture, using a nice group-theoretic description
proposed by Yurke et al. \cite{YMK86}. Using the boson
annihilation operators $a_1$ and $a_2$ of the two input modes,
one constructs the operators
\begin{eqnarray}
  \label{eq:Schwinger}
& & J_x = 
( a_1^{\dagger} a_2 + a_2^{\dagger} a_1 )/2 , \nonumber \\
& & J_y = 
- {\mathrm{i}} ( a_1^{\dagger} a_2 - a_2^{\dagger} a_1 )/2 , \\
& & J_z = 
( a_1^{\dagger} a_1 - a_2^{\dagger} a_2 )/2 . \nonumber
\end{eqnarray}
These operators form the two-boson realization of the su(2)
Lie algebra,
$[ J_k , J_l ] = {\mathrm{i}} \epsilon_{k l m} J_m$.
The Casimir operator is a constant, ${\mathbf{J}}^2 = j (j+1)$, 
for any unitary irreducible representation of the SU(2) group;
so the representations are labeled by a single index $j$ that
takes the values $j = 0,1/2,1,3/2,\ldots$. The representation
Hilbert space $\mathcal{H}_j$ is spanned by the complete 
orthonormal basis $|j,m\rangle$ ($m = j,j-1,\ldots,-j$).
Using Eq.~(\ref{eq:Schwinger}), one finds
\begin{equation}
{\mathbf{J}}^2 = \mbox{$\frac{1}{2}$} N 
\left( \mbox{$\frac{1}{2}$} N + 1 \right) ,
\hspace{8mm}
N = a_1^{\dagger} a_1 + a_2^{\dagger} a_2 ,
\end{equation}
where $N$ is the total number of photons entering the 
interferometer. We see that $N$ is an SU(2) invariant; 
if the input state of the two-mode light field belongs to
$\mathcal{H}_j$, then $N = 2 j$. 

The actions of the interferometer elements on the column-vector
${\mathbf{J}} = ( J_x , J_y , J_z )^T$ can be represented as 
rotations in the 3-dimensional space \cite{YMK86}. The first 
mixing in the beam splitter produces a rotation around the $y$ 
axis by $-\pi/2$, with the transformation matrix
${\mathsf{R}}_y (-\pi/2)$.
The second mixing corresponds to the opposite rotation, with
the transformation matrix ${\mathsf{R}}_y (\pi/2)$. 
The relative phase shift produces a rotation around the $z$ 
axis by $\phi$, with the transformation matrix
${\mathsf{R}}_z (\phi)$.
The overall transformation performed on ${\mathbf{J}}$ is
the rotation by $\phi$ around the $x$ axis,
\begin{equation}
{\mathsf{R}}_x (\phi) = {\mathsf{R}}_y (\pi/2) 
{\mathsf{R}}_z (\phi) {\mathsf{R}}_y (-\pi/2) .
\label{eq:Rx}
\end{equation}

The information on the phase shift $\phi$ is inferred from the
photon statistics of the output beams. Usually, one measures
the difference between the number of photons in the two output 
modes, 
\begin{equation}
q_{{\mathrm{out}}} = 2 J_{z\, {\mathrm{out}}} 
= 2 [ (\sin\phi) J_y + (\cos\phi) J_z ] .
\end{equation} 
If we assume that there are no losses in the interferometer
and the classical sources of noise are well suppressed, then
the uncertainty in the relative position shift $z$ of the end 
mirrors is due to two factors \cite{Caves80,Caves81}. 
The first one is the photon-counting noise. Indeed, since there 
are quantum fluctuations in $q_{{\mathrm{out}}}$, a phase shift is 
detectable only if it induces a change in 
$\langle q_{{\mathrm{out}}} \rangle$ which is larger than the
uncertainty $\Delta q_{{\mathrm{out}}}$. Consequently, the 
uncertainty in the phase shift due to the photon-counting 
noise is 
\begin{equation}
(\Delta \phi)^2_{{\mathrm{pc}}} = 
\frac{ (\Delta q_{{\mathrm{out}}})^2 }{
( \partial \langle q_{{\mathrm{out}}} \rangle / 
\partial \phi )^2 } .
\end{equation}
If the detection is made on a dark fringe ($\phi = \pi/2$
in the unperturbed state), then the contribution of the 
photon-counting noise is 
\begin{equation}
  \label{eq:dz-pc}
(\Delta z)^2_{{\mathrm{pc}}} = A_{{\mathrm{pc}}}
\frac{ ( \Delta J_y )^2 }{  \langle J_z \rangle^2 } ,
\hspace{8mm}
A_{{\mathrm{pc}}} = \left( \frac{L}{\omega \tau} \right)^2 .
\end{equation}
The second source of noise is due to quantum fluctuations in the
radiation pressure. The difference between the momenta transferred
by light to the end mirrors, ${\mathcal{P}} = p_2 - p_1$, is easily
found to be
${\mathcal{P}} = (2 \hbar \omega \tau/ L) J_x$.
The relative shift in the positions of the end mirrors, 
due to the transferred momenta, is $(\tau/m) {\mathcal{P}}$.
Therefore, the contribution of the radiation-pressure noise is 
\begin{equation}
  \label{eq:dz-rp}
(\Delta z)^2_{{\mathrm{rp}}} =  
A_{{\mathrm{rp}}} (2 \Delta J_x )^2 ,
\hspace{8mm} A_{{\mathrm{rp}}} = 
\left( \frac{ \hbar \omega \tau^2 }{m L} \right)^2 .
\end{equation}

Consider the standard case when the coherent laser beam of 
amplitude $\alpha$ enters the interferometer's one input port, while 
the vacuum enters the other. This input state
$|{\mathrm{in}}\rangle = |\alpha\rangle_1 |0\rangle_2$
(where $|0\rangle$ is the vacuum and $|\alpha\rangle = 
\exp(\alpha a^{\dagger} - \alpha^{\ast} a) |0\rangle$ 
is the coherent state), satisfies
\begin{eqnarray*}
& \langle J_x \rangle = \langle J_y \rangle = 0 , \hspace{8mm} 
\langle J_x^2 \rangle = \langle J_y^2 \rangle = |\alpha|^2 /4 , & \\
& \langle J_z \rangle = |\alpha|^2 /2 , \hspace{8mm}
\langle N \rangle \equiv \bar{N} = |\alpha|^2 . &
\end{eqnarray*}
Using these results, one finds 
\begin{equation}
  \label{eq:cohnoise}
(\Delta z)^2 = (\Delta z)^2_{{\mathrm{pc}}} + 
(\Delta z)^2_{{\mathrm{rp}}} = 
A_{{\mathrm{pc}}} \bar{N}^{-1} + A_{{\mathrm{rp}}} \bar{N} . 
\end{equation}
Optimizing $(\Delta z)^2$ as a function of $\bar{N}$, one obtains
\begin{equation}
  \label{eq:Nopt}
\bar{N}_{\mathrm{opt}} = \frac{ m L^2 }{ \hbar \omega^2 \tau^3 } ,
\end{equation}
and $P_{\mathrm{opt}} = \hbar \omega \bar{N}_{\mathrm{opt}}/ \tau$
is given by Eq.~(\ref{eq:Popt}), while the optimum value
of $\Delta z$ is the SQL of Eq.~(\ref{eq:zSQL}).

The characteristic noise behavior of Eq.~(\ref{eq:cohnoise}) is 
sometimes explained by the Poissonian photon statistics of the 
coherent state (i.e., by the fact that 
$\Delta N_1 = \langle N_1 \rangle^{1/2}$, 
with $N_1 = a_1^{\dagger} a_1$). However, it is not difficult
to see that this explanation is principally wrong. 
A simple calculation shows that if instead of the coherent state 
$|\alpha\rangle$ at the first input port we will use an arbitrary 
state (pure or mixture) of the single-mode light field, the same 
result (\ref{eq:cohnoise}) will hold. This will be true, 
regardless of the statistical properties of the photon state at the 
first input port, as long as the vacuum enters the second input 
port.

Caves \cite{Caves81} proposed to reduce the optimum power needed
to achieve the SQL by using the squeezed vacuum in the second 
input port. If the carrier mode entering the first input port
is in the coherent state $|\alpha\rangle$, the two-mode
input state is given by
$|{\mathrm{in}}\rangle = |\alpha\rangle_1 |\xi\rangle_2$,
where $|\xi\rangle = \exp(\frac{1}{2} \xi a^{\dagger 2} - 
\frac{1}{2} \xi^{\ast} a^2 ) |0\rangle$ 
(with $\xi = r {\mathrm{e}}^{ {\mathrm{i}} \theta}$) is the 
squeezed vacuum state.
If one takes $\theta = 0$ and real $\alpha$, the input state
satisfies
\begin{eqnarray*}
& \langle J_x \rangle = \langle J_y \rangle = 0 , \hspace{8mm} 
\langle J_{x,y}^2 \rangle = 
(\alpha^2 {\mathrm{e}}^{\pm 2 r} + \sinh^2 r)/4 , & \\
& \langle J_z \rangle = (\alpha^2 - \sinh^2 r)/2 , 
\hspace{8mm}
\bar{N} = \alpha^2 + \sinh^2 r . &
\end{eqnarray*}
In the usual situation, $\alpha^2 \gg \sinh^2 r$, so one derives
\begin{equation}
  \label{eq:sqnoise}
(\Delta z)^2 \simeq
A_{{\mathrm{pc}}} {\mathrm{e}}^{-2 r} \bar{N}^{-1} + 
A_{{\mathrm{rp}}} {\mathrm{e}}^{2 r} \bar{N} . 
\end{equation}
The use of the squeezed vacuum reduces the
photon-counting noise at the expense of the radiation-pressure
noise. This results in the reduced optimum light power:
\begin{equation}
  \label{eq:Popt-sq}
P_{\mathrm{opt}}(r) \simeq P_{\mathrm{opt}}(r=0)\, 
{\mathrm{e}}^{-2 r} ,
\end{equation}
while the optimum value of $\Delta z$ remains the SQL of 
Eq.~(\ref{eq:zSQL}).
However, it is erroneous to think that the reduction of
the optimum light power is the merit of the squeezed vacuum
alone; actually, the state of the carrier mode is important
as well. While the state in the carrier mode was mixed with
the phase-insensitive vacuum, the behavior of the quantum 
noise was determined by the mean number of carrier photons 
only. However, when mixing the carrier mode with a 
phase-sensitive state (e.g., with the squeezed vacuum), the 
precise matching between the quantum states of the two
modes is important. For example, if the state of the carrier
mode satisfies $\langle a_1^{\dagger 2} + a_1^2 \rangle = 0$
and $\langle N_1 \rangle \gg \sinh^2 r$, then we obtain 
\begin{equation}
(\Delta z)^2 \simeq 
(A_{{\mathrm{pc}}} \bar{N}^{-1} + A_{{\mathrm{rp}}} \bar{N}) 
\cosh 2 r . 
\end{equation}
This results in the same optimum power (\ref{eq:Popt}) as for 
the normal vacuum case, but the sensitivity deteriorates:
$(\Delta z)_{\mathrm{opt}}^2 \simeq 
(\Delta z)_{\mathrm{SQL}}^2 \cosh 2 r$.
For example, this will be the case for the carrier mode in 
the coherent state $|\alpha\rangle$ with $\arg \alpha = \pi/4$ 
or in any phase-insensitive state (a state is called 
phase-insensitive, if its density matrix is diagonal in the 
Fock basis; examples are the Fock states themselves or the 
thermal state). 
On the other hand, if the carrier mode in an arbitrary state
is mixed with a phase-insensitive state (or, more generally,
with any state satisfying $\langle a_2^2 \rangle = 0$), then
\begin{displaymath}
(\Delta z)^2 =
(2 \bar{N}_1 \bar{N}_2 + \bar{N}_1 + \bar{N}_2)
\left[ A_{{\mathrm{pc}}} ( \bar{N}_1 - \bar{N}_2 )^{-2} 
+ A_{{\mathrm{rp}}}  \right] .
\end{displaymath}
Here, we used notation 
$\bar{N}_k = \langle a_k^{\dagger} a_k \rangle$, $k=1,2$.
Clearly, the quantum noise cannot be reduced here, compared
to the vacuum case; in particular, for 
$\bar{N}_1 \gg \bar{N}_2$, the optimum power remains as in
Eq.~(\ref{eq:Popt}), but the sensitivity deteriorates:
$(\Delta z)_{\mathrm{opt}}^2 \simeq 
(\Delta z)_{\mathrm{SQL}}^2 (1 + 2 \bar{N}_2 )$.

From the above arguments, one understands that the reduction
of the optimum power can be achieved with a proper phase
matching between the two input modes. In this relation, it 
is interesting to consider input states which lead to
the Heisenberg-limited photon-counting noise 
\cite{YMK86,HoBu93,SaMi95,HiMl93,BM96,Kim98}. 
It is well known that the shot-noise limit 
$(\Delta \phi)_{{\mathrm{pc}}} = \bar{N}^{-1/2}$, achieved
with the vacuum at the second input port, is not a fundamental
one. Using the uncertainty relation
$(\Delta J_x) (\Delta J_y) \geq \frac{1}{2} 
|\langle J_z \rangle|$, one obtains 
$(\Delta \phi)_{{\mathrm{pc}}} \geq (2 \Delta J_x)^{-1}$.
Since for any input state 
$|{\mathrm{in}}\rangle \in {\mathcal{H}}_j$ the relation
$(\Delta J_x)^2 \leq \frac{1}{2} j (j+1)$ holds, one finds
the Heisenberg limit
\begin{equation}
  \label{eq:Hlimit}
(\Delta \phi)_{{\mathrm{pc}}} \geq [2j (j+1)]^{-1/2} .
\end{equation}
Consequently, for large photon numbers ($\bar{N} = 2 j \gg 1$),
the photon-counting noise $(\Delta z)_{{\mathrm{pc}}}$ scales
as $1/P$.

It can be shown that the shot-noise limit can be surpassed with
the so-called intelligent (minimum-uncertainty) states 
\cite{HiMl93,BM96}. (The use of intelligent states for achieving the 
Heisenberg limit in spectroscopy was discussed in \cite{spectr}.)
The $J_x$-$J_y$ intelligent states, by their definition, 
equalize the uncertainty relation: $(\Delta J_x) (\Delta J_y) 
= \frac{1}{2} |\langle J_z \rangle|$. These states are
determined by the eigenvalue equation
\begin{equation}
  \label{eq:is-eigen}
(\eta J_x - {\mathrm{i}} J_y) |\lambda,\eta\rangle = 
\lambda |\lambda,\eta\rangle .
\end{equation}
The spectrum is discrete: 
$\lambda = {\mathrm{i}} m_0 \sqrt{1-\eta^2}$,
where $m_0 = j,j-1,\ldots,-j$, and $\eta$ is a real parameter
given by $|\eta| = \Delta J_y / \Delta J_x$.
Recently, a method for experimental generation of the SU(2)
intelligent states was proposed in \cite{LuPe96}.

If the two-mode light field entering the interferometer is
prepared in the $J_x$-$J_y$ intelligent state, the quantum 
noise takes the form
\begin{equation}
  \label{eq:intnoise}
(\Delta z)^2 = 
A_{{\mathrm{pc}}} (2 \Delta J_x)^{-2} + 
A_{{\mathrm{rp}}} (2 \Delta J_x)^{2} .
\end{equation}
For $|\eta| < 1$, the intelligent states are squeezed in $J_y$
and anti-squeezed in $J_x$, thereby reducing the photon-counting
noise below the shot-noise limit, on account of increasing
contribution of the radiation-pressure noise.
For $\eta \rightarrow 0$, one obtains \cite{Brif96}
\begin{equation}
(2 \Delta J_x)^2 = 2 | \langle J_z \rangle /\eta | 
\simeq 2 (j^2 - m_0^2 + j) ,
\end{equation}
and the Heisenberg limit for the photon-counting noise is 
achieved when $m_0 = 0$. Then, for large photon numbers 
($\bar{N} = 2 j \gg 1$), we obtain
\begin{equation}
  \label{eq:HLnoise}
(\Delta z)^2 \simeq
2 A_{{\mathrm{pc}}} \bar{N}^{-2} + 
\mbox{$\frac{1}{2}$} A_{{\mathrm{rp}}} \bar{N}^{2} .
\end{equation}
Optimizing $(\Delta z)^2$ as a function of $\bar{N}$, we find 
$(\Delta z)_{\mathrm{opt}} \simeq (\Delta z)_{\mathrm{SQL}}$,
but the optimum light power needed to achieve the SQL is
dramatically reduced:
\begin{equation}
\bar{N}_{\mathrm{opt}} = \left(\frac{2 m L^2}{\hbar \omega^2 \tau^3}
\right)^{1/2} , \hspace{8mm}
P_{\mathrm{opt}} = \left(\frac{2 \hbar m L^2}{\tau^5}
\right)^{1/2} .
\end{equation}
Using the parameters for the initial LIGO, we find the values
$\bar{N}_{\mathrm{opt}} \simeq 4.3 \times 10^{10}$ and 
$P_{\mathrm{opt}} \simeq 9\, \mu\mathrm{W}$.
Compare these values with those obtained in the standard
configuration (with the vacuum at the second port):
$\bar{N}_{\mathrm{opt}} \simeq 9.2 \times 10^{20}$ and 
$P_{\mathrm{opt}} \simeq 191\, \mathrm{kW}$.
We see that the use of the intelligent states with the 
Heisenberg-limited photon-counting noise can in principle reduce 
the optimum light power by a factor $\sim 2 \times 10^{10}$.

There were proposals to achieve the Heisenberg limit for the 
photon-counting noise by driving the interferometer with two 
Fock states containing equal numbers of photons 
\cite{HoBu93,Kim98}. (The use of Fock states was also proposed 
for Heisenberg-limited interferometry with matter waves 
\cite{Dow98} and for Heisenberg-limited spectroscopy with 
degenerate Bose-Einstein gases \cite{BoKa97}.)
The corresponding input state
is $|{\mathrm{in}}\rangle = |n\rangle_1 |n\rangle_2
= |j,0\rangle$ with $j=n=\bar{N}/2$. Clearly, this input
state cannot be used when one measures the photon difference 
$q_{{\mathrm{out}}}$ at the output. However, as was shown
in Ref.\ \cite{Kim98}, the Heisenberg-limited photon-counting 
noise is achieved by measuring the squared difference
$S = q_{{\mathrm{out}}}^2 = 4 J_{z\, {\mathrm{out}}}^2$ 
at the output. (We do not discuss here technical problems 
involved in this kind of measurement.) In this case, the 
uncertainty in the phase shift due to the photon-counting 
noise is
\begin{equation}
(\Delta \phi)_{{\mathrm{pc}}}^2 = \frac{ (\Delta S)^2 }{
( \partial \langle S \rangle / \partial \phi )^2 } 
= \frac{\tan^2 \phi}{8} + 
\frac{ 2 - \tan^2 \phi}{4 j (j+1) } .
\end{equation}
For $\phi = 0$ (this corresponds to a dark fringe for the 
measurement of $S$), the Heisenberg limit is achieved:
$(\Delta \phi)_{\mathrm{pc}} = [2 j (j+1)]^{-1/2}$.
Of course, this improvement is on account of the 
corresponding increase in the radiation-pressure noise,
because $(\Delta J_x)^2 = \frac{1}{2} j (j+1)$ takes
its maximum value. Therefore, for large photon numbers, 
we obtain $(\Delta z)_{{\mathrm{pc}}} \simeq 
(2 A_{{\mathrm{pc}}})^{1/2} /\bar{N}$
and $(\Delta z)_{{\mathrm{rp}}} \simeq 
(A_{{\mathrm{rp}}}/2)^{1/2} \bar{N}$, recovering
the result of Eq.~(\ref{eq:HLnoise}). 
In fact, this quantum noise behavior and the corresponding 
reduction of the optimum power are characteristic for input
states with the Heisenberg-limited photon-counting noise.

It should be emphasized that above results are valid for 
a \emph{lossless} interferometer, while mirrors of realistic 
detectors (for example, LIGO) do have losses. 
The Heisenberg-limited photon-counting noise can be achieved
only if the losses are sufficiently small. Let $\Gamma$ be
the dimensionless coefficient of losses, defined by
$\bar{N}_{{\mathrm{out}}} = \bar{N} {\mathrm{e}}^{- \Gamma}
\simeq \bar{N} (1 - \Gamma)$.
In the case of input Fock states, a simple analysis 
shows \cite{Presk} that the Heisenberg-limited photon-counting 
noise $(\Delta \phi)_{{\mathrm{pc}}} \sim \bar{N}^{-1}$ can be 
obtained only for $\bar{N} \Gamma < 1/2$. Sure, the value 
$\Gamma \sim 10^{-11}$ is impossible to achieve with the present
technology. The problem of losses is of great practical 
importance but nevertheless it does not cross out the principal 
value of the idea to reduce the optimum light power by using 
nonclassical input states. For example, one can imagine a
realization of this idea with a prototype interferometer
which should have smaller $m$ and $L$ and larger $\tau$.

In conclusion, we analyzed the behavior of quantum noise, 
which limits the sensitivity of interferometric 
gravitational-wave detectors, for various input states
of the light field. We found that by using nonclassical
input states exhibiting the Heisenberg-limited photon-counting 
noise, the optimum light power needed to achieve the SQL 
can be significantly reduced, compared
to the usual configuration. Of course, a practical
realization of Heisenberg-limited interferometry will
depend on future theoretical and experimental progress 
in the methods for production of stable and sufficiently 
powerful sources of nonclassical light and on technical
solutions for the reduction of losses and classical sources 
of noise.

The author thanks Kip S. Thorne and John Preskill for stimulating
discussions and Ady Mann and Albert Lazzarini for valuable comments
on the manuscript. Financial support from the Lester Deutsch Fund
is gratefully acknowledged. This work was supported in part by the
Institute of Theoretical Physics at the Department of Physics at
the Technion, and the author is
grateful to the Institute for hospitality during his visit to
the Technion. The LIGO Project is supported by the National 
Science Foundation under the cooperative agreement PHY-9210038.

\end{multicols}

\end{document}